\documentclass[aps,pre,twocolumn,footinbib,superscriptaddress,floatfix]{revtex4}

\usepackage{color}
\usepackage{epsfig}
\usepackage{latexsym}
\usepackage{amssymb}
\usepackage{amsmath}
\usepackage{algorithm}
\usepackage{algorithmic}
\usepackage{latexsym}
\usepackage{fancyhdr}
\usepackage{verbatim}
\usepackage{tabularx}
\usepackage{epsfig}
\usepackage{amsmath}
\usepackage{amssymb}
\usepackage{graphicx}
\usepackage{wasysym}
\usepackage{setspace}

\newcommand{\sset}[1]{ \{#1\} }
\newcommand{\Z}{\mathbf{Z}}

\begin{document}

\title{Strong Resilience of Topological Codes to Depolarization}

\author{H.~Bombin}
\affiliation{Perimeter Institute for Theoretical Physics, Waterloo,
Ontario N2L 2Y5, Canada}

\author{Ruben S.~Andrist}
\affiliation{Theoretische Physik, ETH Zurich, CH-8093 Zurich,
Switzerland}

\author{Masayuki Ohzeki}
\affiliation{Department of Systems Science, Graduate School of
Informatics, Kyoto University, Yoshida-Honmachi, Sakyo-ku, Kyoto
606-8501, Japan}
\affiliation{Dipartimento di Fisica, Universit{\`a} di
Roma `La Sapienza', P.le Aldo Moro 2, 00185, Roma, Italy}

\author{Helmut G.~Katzgraber}
\affiliation {Department of Physics and Astronomy, Texas A\&M
University, College Station, Texas 77843-4242, USA}
\affiliation {Theoretische Physik, ETH Zurich, CH-8093 Zurich,
Switzerland}

\author{M.~A.~Martin-Delgado}
\affiliation{Departamento de F{\'i}sica Te{\'o}rica I, Universidad
Complutense, 28040 Madrid, Spain}

\date{\today}
\begin{abstract}

The inevitable presence of decoherence effects in systems suitable for
quantum computation necessitates effective error-correction schemes to
protect information from noise. We compute the stability of the toric
code to depolarization by mapping the quantum problem onto a classical
disordered eight-vertex Ising model. By studying the stability of the
related ferromagnetic phase both via large-scale Monte Carlo simulations
and via the duality method, we are able to demonstrate an increased
error threshold of 18.9(3)\% when noise correlations are taken into
account.  Remarkably, this agrees within error bars with the result for
a different class of codes---topological color codes---where the mapping
yields interesting new types of interacting eight-vertex models.

\end{abstract}
\maketitle

\section{Introduction}

Moore's law has accurately described the speedup of current
computer technologies for half a century, yet this speedup is
slowly coming to an end due to transistor limitations. A
promising alternative is given by quantum computers. However,
the qubit manipulations required for information processing and
communication are prone to errors because qubits are more sensitive
to noise than their classical counterparts. Consequently, protecting
qubits has become an issue of paramount importance for the success
of quantum computation. The effects of single-qubit operations can
be decomposed into three processes, bit flips, phase flips, as well
as a combination thereof, which can be represented by the three
Pauli matrices $\sigma^x$, $\sigma^z$, and $\sigma^y$, respectively.
This is in contrast to classical bits, which can only suffer from a
single type of error, namely bit flips.

More generally, the notion of a noisy channel is instrumental in
characterizing the disturbing effects on physical qubits. Such a quantum
channel can be described by specifying the probability (or ``qubit error
rate'') $p$ for each of the aforementioned noise types. For instance, if
only $\sigma^x$ occurs, then we have a bit-flip channel.
Here we are interested in channels of the form:
\begin{equation}
	\mathcal{D}_p(\rho) = (1-p)\rho
		+\sum_{w=x,y,z}
		p_w \,\,
		\sigma^w\!\rho\sigma^w\,,
	\label{eq:depol_channel}
\end{equation}
where the density matrix $\rho$ fully describes the quantum state and
the probability for each type of error to occur is $p_{w}\in[0,1]$
with $p:=p_x+p_y+p_z$. The depolarizing channel exhibits equal
probability $p_{w} = p/3$ for each error type to arise. As such, the
depolarizing channel is more general than the bit-flip channel, because
it allows for the unified, \emph{correlated} effect of all three
basic types of errors.  The implications of this error model for the
performance of a quantum code remains an open problem. In addition, the
depolarizing channel plays a fundamental role in quantum-information
protocols where noise has to be taken into account, including
quantum cryptography \cite{shor:00,kraus:05}, quantum distillation
of entanglement \cite{bennett:96}, and even quantum teleportation
\cite{bowen:01}. Experimentally, controllable depolarization has
been realized recently in photonic quantum-information channels
\cite{shaham:11} with a Ti:sapphire pulsed laser and nonlinear
crystals, as well as 2-qubit Bell states \cite{chiuri:11}. Here we
compute the effects of depolarization on a set of entangled qubits.

\subsection{Topological codes}

The goal of quantum error correction \cite{shor:95,steane:96} is to
protect quantum information from decoherence. One approach using
topology is based on encoding (few) \emph{logical} qubits in a
particular state subspace of (typically many) \emph{physical} qubits
which is not disturbed directly by noise. Such a suitable subspace
of states can be defined in terms of a set of commuting observables,
called \emph{check operators},
\begin{equation}
	S^i = \sigma^1\sigma^2\cdots\sigma^{N_i}\,,
\end{equation}
each being a projective measurement with respect to the code subspace
(i.e., the eigenvalue signals errors on participating qubits).
Investigating all stabilizers $S^i$ allows one to limit the set of
possible errors to those compatible with the measured \emph{error
syndrome}. Our best strategy then is to classify the remaining,
nondistinguishable errors according to their effect on the encoded
logical information and undo the effects of the most probable
equivalence class $\overline E$.

A hallmark of \emph{topological} quantum error-correction
codes \cite{kitaev:03,bombin:06,bombin:07,bombin:10,bravyi:10,haah:11} is
the geometrical locality of these check operators: Physical qubits are
placed on a lattice and check operators depend only on a few neighboring
sites. The logical information, which is encoded \emph{globally} in a
subspace of all physical qubits, is preserved as long as we can
successfully detect and correct local errors. If errors on the physical
qubits occur with a probability $p$, the error threshold $p_c$---a key
figure of merit of any quantum code---is defined as the maximum error
probability $p$, such that error classification is achievable. For
error rates larger than $p_c$ the error syndrome's complexity inhibits
unambiguous error recovery. It is therefore of current interest to
find codes where $p_c$ is large.

\subsection{Error threshold as a phase transition}

\begin{figure}
\includegraphics[width=\columnwidth]{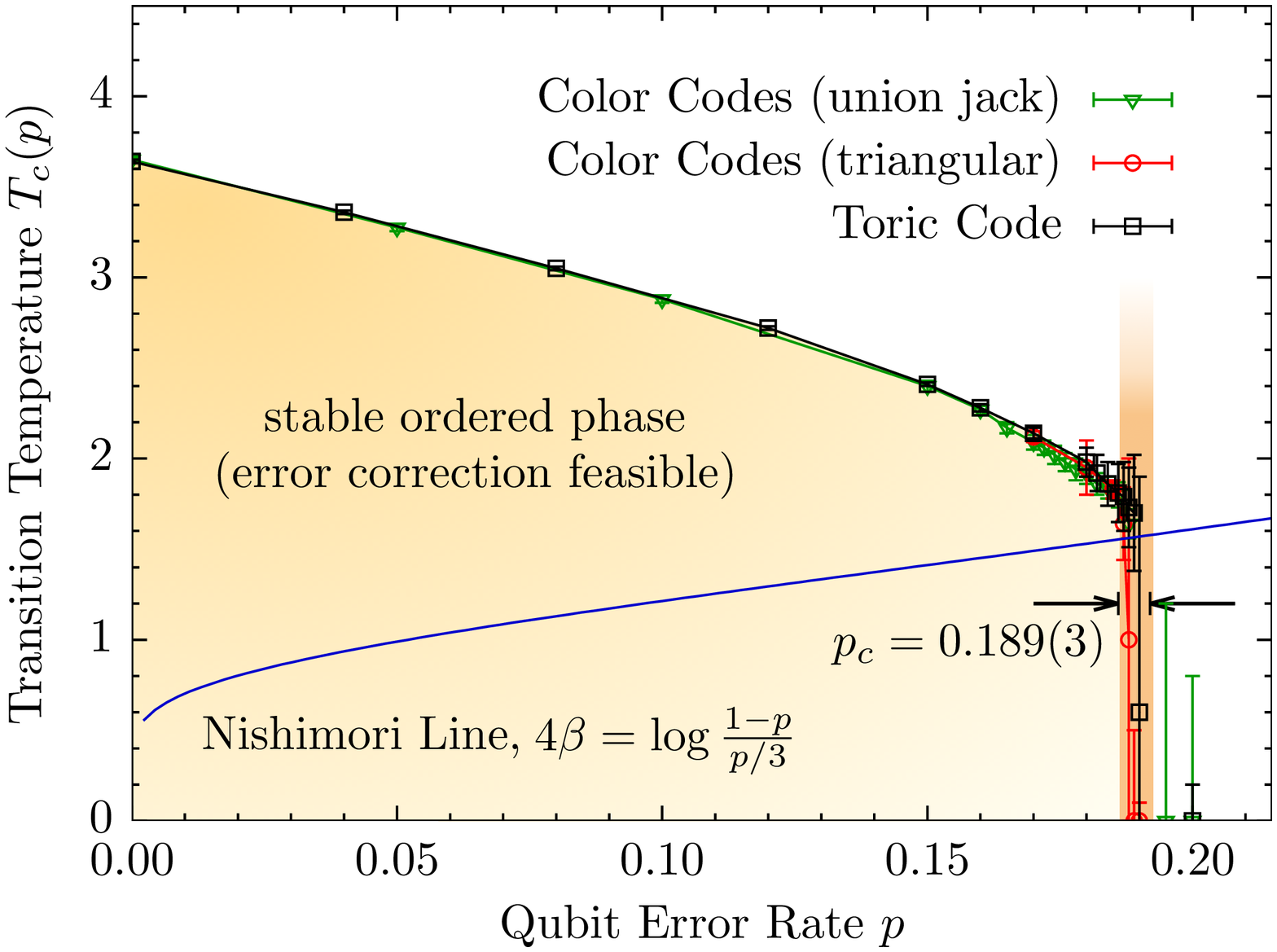}
\caption{(Color online)
Phase boundary estimated from Monte Carlo
simulations for the estimation of the error threshold of the
toric code, as well as two realizations of color codes (see
text). The error threshold $p_c$ corresponds to the point where
the Nishimori lines intersects the phase boundary. Remarkably,
the phase boundaries for all three codes agree within error
bars. The stable ordered phase corresponds to the regime where
quantum error correction is feasible.
}
\label{fig:phasediagram}
\end{figure}

The process of error correction resembles a phase transition and,
indeed, it is possible to connect error correction directly to an
order-disorder phase transition in a suitable classical statistical-mechanical 
model \cite{dennis:02,katzgraber:09c,bombin:10}. One can
derive a Hamiltonian $H_E$ of interacting Ising spins $s_i$, labeled by
a Pauli error $E$ that controls the sign of the couplings, such that the
probability of each equivalence class $\overline E$ is proportional to
the partition function
\begin{equation}
	p(\overline E)\propto Z_E(\beta):= \sum_{\sset{s_i}} 
		{\rm e}^{-\beta H_E(s_i)}\,.
	\label{eq:partition}
\end{equation}
Equation \eqref{eq:partition} has to be interpreted as describing a
random statistical model with quenched couplings and two parameters: the
error probability $p$ governing the fraction of negative interaction
constants $J_{\sigma} \in\{\pm 1\}$, and the inverse temperature $\beta
= 1/T$. For low enough $T$ and $p$ the system orders into a
ferromagnetic state (see Fig.~\ref{fig:phasediagram}). Along the
Nishimori line \cite{nishimori:81} where Eq.~\eqref{eq:partition}
holds, the ordered [disordered] phase corresponds to the topological
code being effective [ineffective]. The intersection of the Nishimori
line and the phase boundary identifies the error threshold $p_c$.

\begin{figure}
\includegraphics[width=\columnwidth]{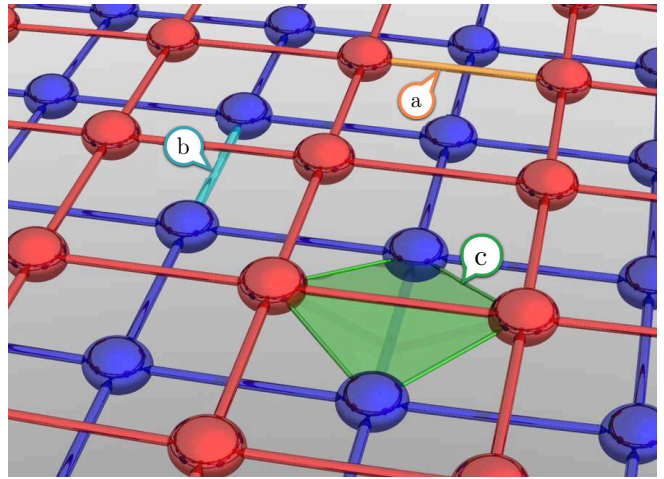}
\caption{(Color online)
When computing the stability of the toric code to
depolarization, the problem maps onto a classical statistical
Ising model on two stacked square lattices. In addition to the
standard two-body interactions for both top (a) and bottom (b)
layers, the resulting Hamiltonian also includes four body
terms (c) that introduce correlations between the layers.
}
\label{fig:toric_lattice}
\end{figure}

The first topological codes studied were toric codes \cite{kitaev:03},
still under intense investigation and scrutiny mainly due to their
simplicity and elegance.  To determine their error threshold, we show
that toric codes under the depolarizing channel connect to the
celebrated eight-vertex model (see Fig.~\ref{fig:toric_lattice}) introduced
by Sutherland \cite{sutherland:70}, as well as Fan and Wu \cite{fan:70},
and whose general solution by Baxter
\cite{baxter:71,baxter:72,baxter:82} stands up as the culmination of a
series of breakthroughs in the theory of phase transitions and critical
phenomena.

The aforementioned mapping onto a statistical-mechanical model
to compute the error tolerance of quantum codes was first applied
to toric codes with bit-flip errors \cite{dennis:02}, connecting
them to the random-bond Ising model. In general, for individual bit
flips the error threshold is $p_c \approx 10.9\%$, and the same is
true for phase flips alone.  Therefore, under depolarizing noise and
correcting separately bit flips and phase flips, the threshold is $p_c'
= (3/2) p_c \approx 16.4\%$.  However, this result neglects
correlations of bit flips and 
phase flips. We estimate the threshold under depolarizing noise for
ideal error correction, such that, in particular, correlations are
taken into account. We find $p_c = 18.9(3)\%$. Remarkably,  the error
threshold increases significantly by taking correlation effects into
account. They should thus not be neglected by recovery algorithms. A
recent advance in this regard is the renormalization approach of Duclos
{\em et~al.}~\cite{duclos:10} where $p_c\simeq 16.4\%$ was confirmed,
still leaving room for further improvement \cite{comment:ref}.
Note also that $p_c$ is very close to the hashing bound $p\simeq
0.1893$ \cite{bennett:96a}, which is also the case for uncorrelated
bit-flip and phase-flip noise \cite{dennis:02,roethlisberger:11}.

\section{Topological stabilizer codes}

\subsection{Error correction in stabilizer codes}

Both toric codes \cite{kitaev:03} and color codes \cite{bombin:06} are
topological stabilizer codes. A stabilizer code is described by a set of
check operators $S_i$ in the Pauli group. That is, they are tensor
products of Pauli operators $\sigma^x$, $\sigma^y$, and $\sigma^z$.
These check operators $S_i$ are a commuting set of observables with
eigenvalue $\pm 1$ that generates an Abelian group $\mathcal S:=\langle
S_i \rangle$ that does not contain $-1$, called the stabilizer group.
Encoded states $|\psi\rangle$ are those for which all check operators
satisfy $S_i|\psi\rangle=+|\psi\rangle$. If errors affect the state,
typically they will change the value of the check operators leaving a
trace that can be used to recover the original state. Note that some
errors are undetectable because they commute with all check operators
and thus leave no trace.

We are interested in noisy channels of the form
\begin{equation}
	\rho_0\longrightarrow \rho_1=\sum_E p(E) \,E\rho_0 E^\dagger\,,
	\label{channel}
\end{equation}
where the sum is over all Pauli group elements $E$, and $p(E)$ denotes
the probability for $E$ to occur. Several different Pauli errors $E$
have the same effect on the encoded state. Therefore, it is convenient
to place them in equivalence classes $\overline E$, such that $E$ is
equivalent to $E'$ when $E\rho E^\dagger=E'\rho{E'}^\dagger$ on an
encoded state $\rho$ or, equivalently, when $EE'$ is proportional to a
product of check operators. Therefore, the total probability for a given
class of errors is given by
\begin{equation}
	p(\overline E)=\sum_{S\in\mathcal S} p(SE)\,.
	\label{prob_class}
\end{equation}
One can choose a set of undetectable errors $D_i$ and use them to
label the error classes compatible with any given syndrome. Namely,
if $E$ is compatible with the syndrome then the possible error classes
are $\overline E$ itself and the classes $\overline {D_iE}$.

The error-correction process starts with the measurement of the check
operators $S_i$. Measuring each $S_i$ yields an eigenvalue $s_i=\pm 1$.
Only certain errors are compatible with these eigenvalues. In
particular, $E$ is compatible with the error syndrome if $ES_i = s_i S_i
E$. Ideally, given a syndrome $s=\sset{s_i}$ one can compute the
relative probabilities $P(\overline E|s)$ of the different error classes
$\overline E$ compatible with $s$. If $\overline {E_s}$ is the class
that maximizes this probability, the best guess is that this is the
error that occurred and thus should be corrected. The net effect of such
an ideal error correction is
\begin{equation}
	\rho_1\longrightarrow \rho_2
		= p_0 \,\rho + \sum_i p_i \,D_i \rho D_i^\dagger\,,
\end{equation}
where the success probability $p_0$ and the probability for an
effective error $D_i$ are
\begin{equation}
	p_0 := \sum_{s} P(\overline E_s)\qquad\qquad p_i
		:= \sum_s p(\overline{D_i E_s})\,.
	\label{success}
\end{equation}
Note that in Eq.~\eqref{success} the sum is over possible
syndromes. Furthermore,
\begin{equation}\label{bound_prob}
	\frac 1 D \leq \frac {P(\overline E_s)}{P(s)} \leq 1
		\qquad\qquad \frac 1 D \leq p_0 \leq 1\,,
\end{equation}
where $D$ is the number of error classes per error syndrome. In
practice this ideal error correction might be too costly from a
computational perspective. Therefore, approximations are needed.

\subsection{Toric codes and color codes}

\begin{figure}
\includegraphics[width=\columnwidth]{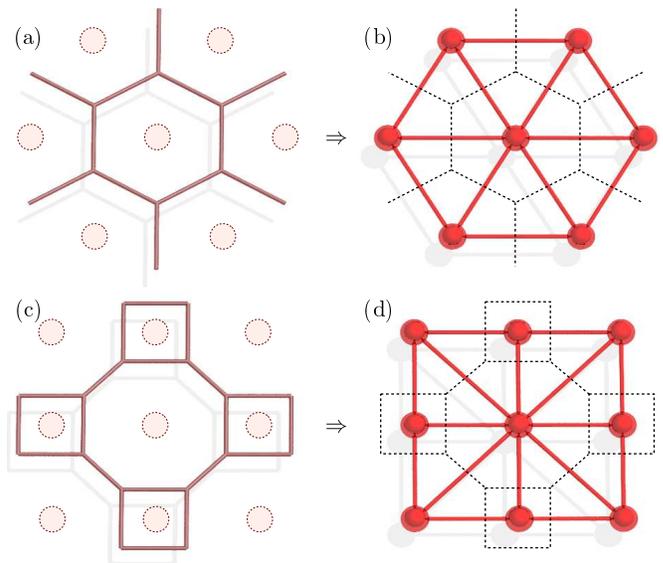}
\caption{(Color online)
For the hexagonal arrangement, there is a stabilizer operator $Z^6$ for
each of the hexagon plaquettes (top left). In the mapping, these
stabilizer operators translate to classical Ising spins, which are
placed on the dual lattice (regular triangular lattice, top right). The
square-octagonal setup (bottom left) has wider computing capabilities
because it allows for a larger class of quantum gates to be implemented.
There are stabilizers $Z^4$ [$Z^8$] on the rectangles [octagons]. The
corresponding dual lattice in the mapping is the union jack lattice
(bottom right).}
\label{fig:mapping}
\end{figure}

Topological codes have two interesting features: First, they can be
defined for different system sizes in such a way that check operators
remain local---involving only a few neighboring qubits---and at the same
time nontrivial undetectable errors are global and thus involve a number
of qubits that depend on the system size. Second, they exhibit an error
threshold. For error rates below the threshold, the success probability
[Eq.~\eqref{success}] approaches $1$ for increasing system size,
whereas $1-p_0$ decreases exponentially.

In toric codes \cite{kitaev:03} physical qubits are placed on the
edges of a square lattice. Notice that for each edge in the direct
lattice there is an edge in the dual lattice. Check operators $S_f$
are attached to faces $f$, either in the direct or the dual lattices.
Toric codes can thus be defined in two similar, but distinct ways:
In the original definition by Kitaev, if $f$ is a face in the direct
[dual] lattice composed by the edges $r$, $s$, $t$ and $u$, then the
corresponding check operator is $S_f:= \sigma^x_r \otimes \sigma^x_s
\otimes \sigma^x_t \otimes \sigma^x_u$ [$S_f:= \sigma^z_r \otimes
\sigma^z_s \otimes \sigma^z_t \otimes \sigma^z_u$]. The second
definition is due to Wen \cite{wen:03} and it does not distinguish
between dual and direct faces. If $f$ has a top edge $r$, a bottom edge
$s$ and side edges $t$, $u$, then we take $S_f:= \sigma^z_r \otimes
\sigma^z_s \otimes \sigma^x_t \otimes \sigma^x_u$. Both definitions are
equivalent up to a rotation of half of the qubits. However, for the
depolarizing channel Kitaev's definition is related to the alternating
eight-vertex model and Wen's definition to the standard eight-vertex model.

In color codes \cite{bombin:06} physical qubits are placed on the
vertices of a trivalent lattice with three-colorable faces, such as, for
example, the honeycomb lattice. There are two check operators $S^x_f$
and $S^z_f$ attached to each face $f$ taking the form
$S^x_f:=\bigotimes_{i} \sigma^x_{i}$ and $S^z_f:=\bigotimes_{i}
\sigma^z_{i}$, respectively, with $i$ running over the qubits on the
vertices of $f$.

Because the computing capabilities of color codes depend on the
underlying lattice where the qubits are placed, we study two different
scenarios: the honeycomb lattice for its simplicity, and a lattice of
octagons and squares that allows for the implementation of additional
types of quantum gates. In the mapping onto a statistical-mechanical
model to compute the error threshold these two arrangements correspond
to the triangular and union jack lattices, respectively (see
Fig.~\ref{fig:mapping}).

\section{Random eight-vertex Ising Models}

To determine the error threshold, we show that topological codes under
the depolarizing channel connect to certain random classical spin
models.

For the toric code, the error-correction process maps onto a
statistical-mechanical \emph{interacting eight-vertex model}
\cite{sutherland:70,fan:70,baxter:71,baxter:72}. Remarkably, this class
of models exhibit critical exponents that depend on the coupling
constants, Eq.~\eqref{Hamiltonian_toric}, thus challenging the very
notion of universality.  Eight-vertex models were originally formulated in
the `electric picture' where the degrees of freedom are electric dipoles
placed at the bonds surrounding each vertex of a square lattice
\cite{baxter:82}, i.e., the number of independent dipole configurations
per vertex is eight. In addition, a mapping to a `magnetic picture' was
found by Wu \cite{wu:71}, as well as Kadanoff and Wegner
\cite{kadanoff:71}: Consider two independent Ising systems, each on a
square lattice, with classical spin variables $s_i$ and $s_k'$ taking on
values $\pm1$, and bonds $J_{ij}$ and $J_{k\ell}'$, respectively.  The
lattices are stacked as shown in Fig.~\ref{fig:toric_lattice} such that
the vertices (spin sites) of one lattice are at the center of the
plaquettes of the other.  The Hamiltonian takes the explicit form
\begin{equation}
	H= -\sum_{+} (J_{ij} s_is_j + J_{k\ell}'
	s_k's_\ell'+J_{+}s_is_js_k's_\ell')\,.
	\label{eq:toric_hamiltonian}
\end{equation}
This can be thought of as two interacting Ising models by means of a
four-spin interaction (denoted by the symbol $+$) between original and
dual lattices.

In fact, two types of eight-vertex models can be related to error correction
in toric codes: the standard eight-vertex model where $J_{ij}=J$
[$J_{ij}=J'$] if a bond is a horizontal [vertical] link, see
Figs.~\ref{fig:toric_lattice}(a) and \ref{fig:toric_lattice}(b); and the
alternating eight-vertex model where $J_{ij}=J$ [$J_{ij}=J'$] if a bond
belongs to the direct [dual] lattice. In both cases, we set the four-spin
interaction to $J_+=J_4$, as depicted in
Fig.~\ref{fig:toric_lattice}(c). Thus, both types of eight-vertex models
share the same lattice structure but differ in the pattern of coupling
constants. This difference has fundamental consequences in the exact
solvability of the model: while the standard eight-vertex model is exactly
solved for arbitrary couplings, the alternating eight-vertex model is
generally not exactly solvable. In fact, the latter corresponds to the
Ashkin-Teller model \cite{ashkin:64}. Notice that when $J_4=0$, the
eight-vertex model reduces to two decoupled Ising models, while for $J_4\neq
0$ the model has two critical temperatures.

The error threshold for correction in quantum codes corresponds
to the critical line separating ordered from disordered phases.
The former represents a situation where quantum error correction can
be performed with arbitrary precision.  Determining the location
of this critical line in eight-vertex models is facilitated by the
existence of a self-duality symmetry in the partition function:
a duality transformation relating a high-temperature eight-vertex model
to a low-temperature one on the same lattice. Self-duality implies
that the coupling constants for 2-spin interactions are isotropic,
i.e., $J=J'$.  Altogether, an isotropic self-dual eight-vertex model has
a critical line given by \cite{baxter:82}:
\begin{equation}
	J' = J\, , \quad\quad\quad {\rm e}^{-2\beta J_4} = \sinh(2\beta J),
	\label{sup:eq:self-duality}
\end{equation}
with the restriction that $J_4\leq J$. The point in the plane
$(J_4=J_c,J=J_c)$ at which the self-dual line ceases to be critical is
given by
\begin{equation}
	\beta J_c = \frac{1}{4} \log (3) \approx 0.2746\ldots .
	\label{critical_point}
\end{equation}
This is already a remarkable and encouraging result because it yields a
critical point which is approximately $60\%$ larger than in the standard
square-lattice two-dimensional Ising model. Note that the error
threshold for bit-flip or phase-flip errors in the Kitaev model is
computed via a mapping to the aforementioned two-dimensional Ising
model. In that case, the critical point can be computed from the
relationship $\sinh(2\beta J_c)=1$, i.e., $\beta J_c=0.4406$. Recall that
the critical exponents depend continuously on the value of $J_4$.

In this work we extend the standard eight-vertex model by adding quenched
disorder to the couplings between the spins. Given that for the
eight-vertex model $\beta J_c \approx 0.2746 $ is smaller than for the
square-lattice Ising model, we can expect to find a larger error
threshold for the depolarizing channel than for bit-flip or phase-flip
errors.

In addition to depolarizing errors in the toric code, where the problem
map onto Eq.~\ref{eq:toric_hamiltonian}, we also study color codes, see
Fig.~\ref{fig:mapping}. In this case the underlying statistical-mechanical 
model to study the error stability to depolarizing errors is
defined on a triangular lattice. There are two Ising variables---$s_i$
and $s_i'$---per site. For convenience, we introduce an artificial third
variable $s_i''=s_i^{\,}s_i'$. The Hamiltonian is then given by:
\begin{equation}
	H_2 = - \sum_{\langle i,j,k\rangle}
		\left(
		J_{ijk} \,s_is_js_k +
		J'_{ijk}\,s_i's_j's_k' +
		J''_{ijk}\,s_i''s_j''s_k''
		\right)\,.
	\label{eq:color_hamiltonian}
\end{equation}
Equation \eqref{eq:color_hamiltonian} is illustrated in
Fig.~\ref{fig:colorcode_lattice} where the top [bottom] layer
corresponds to the $s_i$ [$s_i'$] Ising variables with the corresponding
three-spin interaction term as shown in
Fig.~\ref{fig:colorcode_lattice}(a) [\ref{fig:colorcode_lattice}(b)].
The third term in the Hamiltonian with six-spin interactions is
represented by Fig.~\ref{fig:colorcode_lattice}(c).

When $J_{ijk}'' = 0$ in Eq.~\eqref{eq:color_hamiltonian}, we obtain two
independent triangular 3-body Ising models. Interestingly, this model
can be mapped onto a eight-vertex model on a Kagom\'e lattice
\cite{baxter:78}. Therefore, the color code Hamiltonian $H_2$ in
Eq.~\eqref{eq:color_hamiltonian} can be thought of as an interacting
eight-vertex model (or coupled eight-vertex models). In this work we consider
two different lattice geometries, triangular and union jack (see
Fig.~\ref{fig:mapping}).

\section{Mapping}

\subsection{Spin models for depolarizing noise}

The goal is to relate the stability of a topological stabilizer code to
depolarizing noise to the ordered phase of a suitably chosen classical
spin model. However, here we consider the more general qubit channel
shown in Eq.~\eqref{eq:depol_channel}. This adds transparency to the
mapping and reveals the differences between Kitaev's and Wen's versions
of the toric code with respect to error correction. When
Eq.~\eqref{eq:depol_channel} is applied to each qubit in a code, the net
result is a channel of the form presented in Eq.~\eqref{channel}.  In
particular, the probability for each Pauli error is
\begin{equation}
	p(E) =  (1-p)^n \prod_{w=x,y,z}\left(\frac{p_w}{1-p}\right)^{E_w}\,,
	\label{eq:probability}
\end{equation}
where $n$ is the total number of qubits and $E_w$ the number of
appearances of $\sigma^w$ in the tensor product forming $E$.

The classical spin Hamiltonian is constructed as follows:
\begin{enumerate}\itemsep-1pt

\item{Attach a classical spin $s_i$ to each check operator $S^{i}$.}

\item{Associate with each single-qubit Pauli operator $\sigma$ an
interaction term $J_\sigma s_1^\sigma s_2^\sigma\cdots
s_{N_\sigma}^\sigma$ such that the spins $s_i$ correspond to the check
operators $S^{i}$ affected by $\sigma$, i.e., such that
$S_i\sigma=-\sigma S_i$.}

\item{Attach to each coupling a sign $\tau_\sigma=\pm 1$ dictated by the
Pauli error $E$ labeling the Hamiltonian, through the conditions $\sigma
E=\tau_\sigma E\sigma$.}

\end{enumerate}
The resulting Hamiltonian takes the general form
\begin{equation}
	H_E = - \sum_{\sigma} J_\sigma \,\tau_\sigma\,
		s^\sigma_1s^\sigma_2\cdots s^\sigma_{N_{\sigma}}\,,
	\label{gen_Hamiltonian}
\end{equation}
where the sum is over all Pauli operators $\sigma$ and there are only
three different couplings $J_\sigma$ since we set $J_{\sigma_k^w}:=J_w$,
with $w=x,y,z$ and $k$ the qubit label.

For the mapping, we require the interaction constants to be
\begin{equation}
	J_w = -\frac{1}{4\beta}\log\frac{p_xp_yp_z}{p_w^2(1-p)}
	\,,\quad w=x,y,z\,.
	\label{eq:nishimori_line}
\end{equation}
This relates the error probability in Eq.~\eqref{eq:probability} to the
Boltzmann factor for the ordered state, $\sset{s_i=1}$, given the
interactions generated by $E$:
\begin{equation}
	p(E) \propto {\rm e}^{-\beta H_E( \sset{s_i=1} ) }\,.
\end{equation}
Note that, to recover Eq.~\eqref{eq:partition} just notice that
replacing the error $E \to E' = S^jE$ is equivalent to considering a
different spin configuration in the original Hamiltonian:
\begin{equation}
	H_{S^jE}(\sset{s_i}) = H_E(\sset{(1-2\delta_{ij})s_i})\,.
\end{equation}
Finally, to complete the mapping, the label $E$ in the Hamiltonian must
describe quenched randomness. In particular, the coupling configuration
dictated by $E$ appears with probability $p(E)$. Equivalently, this
means that for every qubit $k$ the probability
$p(\tau_k^x,\tau_k^y,\tau_k^z)$ for each configuration of coupling signs
is given by
\begin{eqnarray}
p(+1,+1,+1) &=& 1-p,	\\ \nonumber
p(+1,-1,-1) &=& p_x,	\\ \nonumber
p(-1,+1,-1) &=& p_y,	\\ \nonumber
p(-1,-1,+1) &=& p_z.	\\ \nonumber
\end{eqnarray}
In the case of the depolarizing channel
\begin{equation}
p_x=p_y=p_z=p/3 \quad {\rm and}\quad J_x=J_y=J_z=J .
\end{equation} 
The resulting model has two parameters, $p$ and $\beta J$ with $\beta J =
1/T$. For low $p$ and $T$ the model orders ferromagnetically and along
the Nishimori line,
\begin{equation}
	{\rm e}^{-4\beta J} = \frac{p/3}{(1-p)} ;
	\label{Nishimori}
\end{equation}
this order is equivalent to the noise being correctable. Indeed,
comparing error-class probabilities amounts to computing free-energy
differences
\begin{equation}
	\frac {P(\overline{D_i E})}{P(\overline E)}
	= \frac{Z_{D_iE}(\beta)}{Z_{E}(\beta)}
	=: {\rm e}^{-\beta \Delta_i (\beta, E)}\,.
\end{equation}

In topological codes we expect the existence of an error threshold
$p_c$---or several for different error types, but we do not need such
generality. When $p<p_c$ in the limit of large systems the success
probability is expected to approach unity, i.e., $p_0\longrightarrow 1$
and thus due to Eq.~\eqref{bound_prob} along the Nishimori line the
free-energy difference is asymptotically infinite, because for any real
$t$, $P(\Delta_i (\beta, E)>t)\longrightarrow 1$.  Similarly, when
$p>p_c$, the success probability is expected to become minimal
($p_0\longrightarrow 1/D$) and thus the free-energy difference converges
in probability zero, so that for any $t>0$ we have $P(|\Delta_i (\beta,
E)|<t)\longrightarrow 1$.  This shows that the-free energy differences
$\Delta_i$ are order parameters and $p_c$ is the critical value of $p$ along
the Nishimori line. In the models of interest here, these are
domain-wall free energies.

\subsection{Models for toric and color codes}

\begin{figure}
\includegraphics[width=\columnwidth]{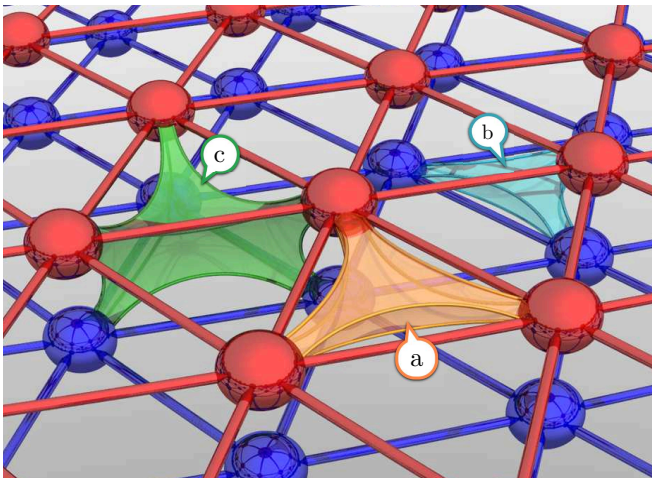}
\caption{(Color online)
For topological color codes, qubits are arranged
on trivalent lattices (hexagonal or square-octagonal). These
codes are mapped to triangular lattices (triangular and union
jack, respectively) with plaquette interactions (a,b) on each
layer, as well as six-body interactions correlating the two
layers (c).
}
\label{fig:colorcode_lattice}
\end{figure}

Let us now study what the above mapping, Eq.~\eqref{gen_Hamiltonian},
produces when applied to toric codes and to topological color codes.

For the toric code, the single-qubit operators $\sigma^x$ and $\sigma^z$
produce 2-body interactions, because each bit flip [phase flip] affects
the stabilizer operators on two 
on two neighboring dual [direct] faces.
The $\sigma^y$ operators, which combine correlated spin-flip and
phase-flip errors, introduce four-body interactions, see
Fig.~\ref{fig:toric_lattice}.  The result is an alternating eight-vertex model
with coupling signs that are parametrized by a Pauli error $E$, namely,
\begin{equation}
	H_E = - \sum_+ (J_x\, \tau_+^x\, s_i s_j
				 + J_z \,\tau_+^z\, s_k' s_l'
				 + J_y \,\tau_+^y\, s_i s_j s_k' s_l')\,.
	\label{Hamiltonian_toric}
\end{equation}
The classical spin variables $s_i$ [$s_k'$] live on the top [bottom]
layer of two stacked two-dimensional Ising square lattices with
interaction constants $J_x$ [$J_z$] (see Fig.~\ref{fig:toric_lattice}).
The two layers are shifted by half a lattice spacing and the third term
in Eq.~\eqref{Hamiltonian_toric} describes the combined four spin
interaction at each of the crossings ``$+$''.  Note that in
Eq.~\eqref{Hamiltonian_toric} there is one qubit per cross $+$. For
Wen's toric code one recovers the standard eight-vertex model. In either
case, there is a global $\Z_2 \times \Z_2$ symmetry because one can flip
all spins in each lattice separately without affecting the total energy.

In the case of color codes there is one spin per face. The $\sigma^x$
and $\sigma^z$ single-qubit operators produce 3-body interactions in
Eq.~\eqref{gen_Hamiltonian}, whereas $\sigma^y$ operators produce 6-body
interactions. The Hamiltonian is then given by
Eq.~\eqref{eq:color_hamiltonian} but with coupling signs parametrized by
a Pauli error $E$, namely,
\begin{equation}
	H_E  = - \sum_{\langle i,j,k\rangle } \sum_{w=x,y,z} J_w\,
		\tau_{ijk}^w\, s_i^w s_j^w s_k^w\,,
	\label{Hamiltonian_cc}
\end{equation}
with $s_i^x s_i^y s_i^z=1$. Therefore, we obtain two stacked triangular
lattices having three and six-body interactions (see
Fig.~\ref{fig:colorcode_lattice}), with the six-body interactions
introducing correlations between both layers.  In this case the global
symmetry is $\Z_2 \times\Z_2 \times\Z_2 \times \Z_2$. Indeed, the sites
can be colored with three colors in such a way that each triangles has a
site of each color. Thus one can flip all spins for two given colors in
either of the two lattices separately without affecting the total
energy.

For $p=0$ in Eqs.~\eqref{Hamiltonian_toric} and \eqref{Hamiltonian_cc}
self-duality predicts a critical temperature of $T_c = 1/\beta J_c
\simeq 3.641$, a value that we confirm numerically in our Monte Carlo
simulations.

\section{Monte Carlo Simulations}

We investigate the classical statistical spin models acquired in the
mapping, Eq.~\eqref{Hamiltonian_toric} and Eq.~\eqref{Hamiltonian_cc},
via large-scale classical Monte Carlo simulations using the parallel
tempering Monte Carlo technique \cite{hukushima:96}.

In the parallel tempering Monte Carlo method, several identical
copies of the system at different temperatures are simulated. In
addition to local simple Monte Carlo (Metropolis) spin updates
\cite{newman:99}, one performs global moves in which the temperatures
of two neighboring copies are exchanged. It is important to
select the position of the individual temperatures carefully such
that the acceptance probabilities for the global moves are large
enough \cite{katzgraber:06} and each copy performs a random walk in
temperature space.  This, in turn, allows each copy to efficiently
sample the rough energy landscape, therefore speeding up the simulation
enormously.

Detecting the transition temperature $T_c(p)$ for different fixed
amounts of disorder allows us to pinpoint the phase boundary in the
$p\,$---$T$ phase diagram. The error threshold $p_c$ is then given
by the intersection of the phase boundary with the Nishimori line.

\subsection{Observables and Simulation Details}

For the toric code, it is expedient to partition the lattice
into two sublattices such that the only interconnection is
given by the four-body-interactions of the Hamiltonian in
Eq.~\eqref{Hamiltonian_toric}. The ground state of the pure system
is realized when the spins of each sublattice are aligned (but the
alignment may be different as the sign would cancel out in both the
two and four-spin terms). In this case the sublattice magnetization
is a good order parameter,
\begin{equation}
	m = \frac{1}{N_\mathcal{P}}\sum_{i\in\mathcal{P}} S_i\,,
	\label{eq:order_parameter}
\end{equation}
where $\mathcal P$ denotes one of the sublattices. Similarly, we note
that each layer of the triangular lattice for color codes is tripartite
with spins aligned in each sublattice for all realizations of the pure
system's ground state. Hence, we can define an order parameter
analogous to Eq.~\eqref{eq:order_parameter} for a suitable
sublattice $\mathcal{P}'$. Note that particular caution is required when
implementing the periodic boundary conditions to ensure that these
distinct sublattices are well defined. We can now use the magnetization
defined in Eq.~\eqref{eq:order_parameter} to construct the 
wave-vector-dependent magnetic susceptibility,
\begin{equation}
	\chi_m(\mathbf{k}) =
	\frac{1}{N_\mathcal{P}}
	\left\langle\left(
	\sum_{i\in\mathcal{P}} S_i
	{\rm e}^{i\mathbf{k}\cdot\mathbf{R}_i}
	\right)^2\right\rangle_T \, ,
	\label{eq:susceptibility}
\end{equation}
where $\langle\cdots\rangle_T$ denotes a thermal average and
$\mathbf{R}_i$ is the spatial location of the spin $S_i$. From 
Eq.~\eqref{eq:susceptibility} we construct the two-point finite-size
correlation function \cite{palassini:99b},
\begin{equation}
	\xi_L = \frac{1}{2\sin(k_{\rm min}/2)}\sqrt{
		\frac{[\chi_m(\mathbf{0})]_{\rm av}}
		{[\chi_m(\mathbf{k}_{\rm min})]_{\rm av}}-1
	}\,,
	\label{eq:correlation_function}
\end{equation}
where $[\cdots]_{\rm av}$ denotes an average over disorder and
$\mathbf{k}_{\rm min} = (2\pi/L,0,0)$ is the smallest nonzero wave
vector.  Near the transition, $\xi_L$ is expected to scale as
\begin{equation}
	\xi_L/L \sim \tilde X[L^{1/\nu}(T-T_c)]\,,
	\label{eq:xiL_scaling}
\end{equation}
where $\tilde X$ is a dimensionless scaling function. Because at the
transition temperature, $T=T_c$, the argument of
Eq.~\eqref{eq:xiL_scaling} becomes zero (and hence independent of~$L$),
we expect lines of different system sizes to cross at this point. If
however the lines do not meet, we know that no transition occurs in the
studied temperature range.

In all simulations, equilibration is tested using a logarithmic binning
of the data: This is done by verifying that all observables averaged over
logarithmically-increasing amounts of Monte Carlo time are, on average,
time independent. Once the data for all observables agree for three
logarithmic bins we deem the Monte Carlo simulation for that system size
to be in thermal equilibrium. The simulation parameters can be found in
table \ref{tab:simparams}.

\begin{table}[!tb]
\caption{
Simulation parameters: $L$ is the linear system size, $N_{\rm sa}$
is the number of disorder samples, $t_{\rm eq} = 2^{b}$ is the number
of equilibration sweeps (system size times number of single-spin Monte
Carlo updates), $T_{\rm min}$ [$T_{\rm max}$] is the lowest [highest]
temperature, and $N_{\rm T}$ the number of temperatures used.  For the
toric code, we use $L=\{12,16,18,24,32,36\}$, while for color codes
$L=\{12,15,18,24,30,36\}$ following the coloring constraints that
the system size must be a multiple of $3$.
}
\label{tab:simparams}
\vspace*{1mm}
\centering
{\footnotesize
\begin{tabular*}{8 cm}{@{\extracolsep{\fill}} l r r r r r r}
\hline
\hline
$p$ & $L$ & $N_{\rm sa}$ & $b$ & $T_{\rm min}$ & $T_{\rm max}$ &$N_{\rm T}$ \\
\hline
$0.00$        & $12-16$ & $5\,000$  & $18$ & $3.500$ & $4.000$ & $42$\\
$0.00$        & $18-24$ & $1\,000$  & $19$ & $3.500$ & $4.000$ & $42$\\
$0.00$        & $30-36$ & $500$     & $20$ & $3.500$ & $4.000$ & $42$\\
$0.04-0.05$   & $12-16$ & $5\,000$  & $20$ & $3.200$ & $3.800$ & $42$\\
$0.04-0.05$   & $18-24$ & $1\,000$  & $21$ & $3.200$ & $3.800$ & $42$\\
$0.04-0.05$   & $30-36$ & $500$     & $22$ & $3.200$ & $3.800$ & $42$\\
$0.08-0.12$   & $12-16$ & $5\,000$  & $20$ & $2.700$ & $3.500$ & $42$\\
$0.08-0.12$   & $18-24$ & $1\,000$  & $22$ & $2.700$ & $3.500$ & $42$\\
$0.08-0.12$   & $30-36$ & $500$     & $24$ & $2.700$ & $3.500$ & $42$\\
$0.15$        & $12-16$ & $5\,000$  & $20$ & $2.300$ & $3.200$ & $42$\\
$0.15$        & $18-24$ & $1\,000$  & $22$ & $2.300$ & $3.200$ & $42$\\
$0.15$        & $30-36$ & $500$     & $24$ & $2.300$ & $3.200$ & $42$\\
$0.17-0.20$   & $12-16$ & $5\,000$  & $21$ & $1.500$ & $2.800$ & $42$\\
$0.17-0.20$   & $18-24$ & $1\,000$  & $23$ & $1.500$ & $2.800$ & $42$\\
$0.17-0.20$   & $30-36$ & $500$     & $25$ & $1.500$ & $2.800$ & $42$\\
\hline
\hline
\end{tabular*}
}
\end{table}

\subsection{Sample results}

\begin{figure}
\includegraphics[width=\columnwidth]{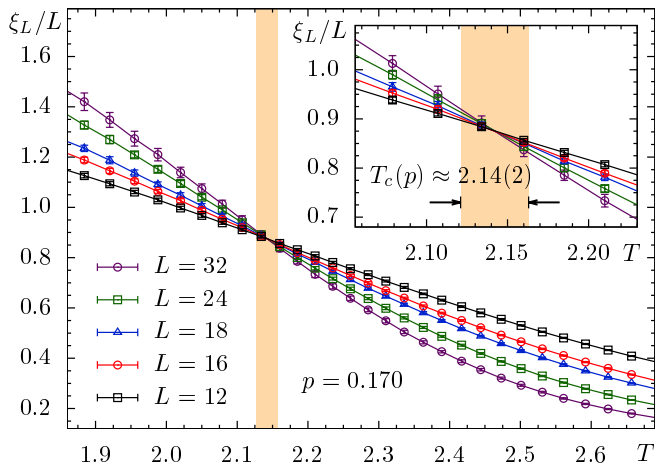}
\caption{(Color online)
Crossing of the correlation function $\xi_L/L$ for the toric code with a
disorder rate of $p=0.170$. The data exhibit a clear crossing at a
transition temperature of $T_c(p)\approx 2.14(2)$. Corrections to
scaling are still minimal at this disorder rate, but increase closer to
the error threshold. Inset: closeup of the area where the crossing
occurs. The conservative estimate for the transition temperature is
indicated by the vertical shade.
}
\label{sup:fig:crossing}
\end{figure}

For the pure system ($p=0$) there is a sharp transition visible directly
in the sublattice magnetization. The transition temperature $T_{c,{\rm
pure}}\approx 3.64$ coincides with the value found analytically. For
larger amounts of disorder, a transition can still be located
precisely by means of the crossings in the two-point finite-size
correlation function [Eq.~\eqref{eq:correlation_function}] for
different system sizes. Sample data for a disorder strength of $p=0.170$
(i.e., this would mean that on average 17\% of the physical qubits are
``broken'') are shown in Fig.~\ref{sup:fig:crossing}, indicating a
transition temperature of $T_c(p) = 2.14(2)$. The error bars are
calculated using a bootstrap analysis of 500 samples. There are
small finite-size effects which are addressed by analyzing the
intersection $T_c^*(L,2L)$ of pairs of system sizes. We estimate the
limit value for $L\to\infty$ by means of a linear fit in a $1/L$-plot --
this is our estimate for the best value in the physically-relevant
thermodynamic limit. For disorder rates approaching the error threshold,
corrections to scaling increase and a careful finite-size scaling
analysis has to be performed to determine $T_c$ \cite{andrist:10}. At
$p=0.189$, the lines only touch marginally such that both the scenario
of a crossing as well as no transition are compatible within error bars.
This gives rise to the large error bars in the phase diagram
(Fig.~\ref{fig:phasediagram}).  For error rates $p>p_c$, the lines do
not meet, indicating that there is no transition in the temperature
range studied.

The crossing of the critical line $T_c(p)$ with the Nishimori line
[Eq.~\eqref{Nishimori}] determines the error threshold to
depolarization. Our (conservative) estimate is $p_c = 0.189(3)$. Our
results are summarized in Fig.~\ref{fig:phasediagram}, which shows the
estimated phase diagram.

\section{Duality method}
\label{sec:duality}

An alternative approach to estimate the critical value $p_c$ is to use
the duality method \cite{ohzeki:09}, originally developed within the
context of spin glasses.

\begin{figure}
\includegraphics[width=\columnwidth]{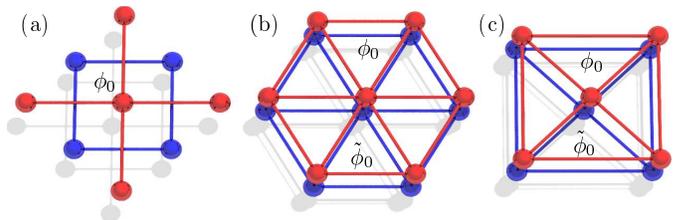}
\caption{(Color online)
Clusters used to estimate the error threshold for the depolarizing
channel.  The blue lines and triangles denote quenched random variables
$\tau^Z$, and the red lines and triangles correspond to $\tau^X$. The
central site is the spin variables summed over. The outer sites
represent the spin variables fixed in the up direction.
}
\label{fig:duality}
\end{figure}

The critical point of a statistical model expressed only by local
interactions between degrees of freedom can be analyzed using the
duality method under the assumption of a unique transition temperature.
The partition function $Z[A]$ is then given by the local Boltzmann
factor $A_{\phi} = \exp(\beta J\cos\pi\phi)$, where $\phi\in \{0,1\}$ is
the difference between adjacent spins such as $\cos(\pi \phi) = \pm 1$.
We define the {\em principal Boltzmann factor} $A_0$ as the case where
all spins are parallel. The partition function has to be invariant under
a Fourier transform, i.e., $Z[A] = Z[A^*]$, where $A^*$ is a dual
principal Boltzmann factor (via a Fourier transformation). In that case
the critical point is determined via the equality $A_0 = A_0^*$. This
implies that all the components of the local Boltzmann factors are equal
for several self-dual models such as the standard Ising model. Although
this self-duality does not work a priori for systems with quenched
disorder in the general case, the method can be applied in a special
subspace called the Nishimori line \cite{ohzeki:09}. The results can be
improved by considering extended local Boltzmann factors over a
restricted range of interactions \cite{ohzeki:09,ohzeki:09a} (see
Fig.~\ref{fig:duality} which illustrates the used clusters). Because the
resulting statistical-mechanical Hamiltonians for both the toric code
and topological color codes are self dual, we can apply this efficient
technique to obtain estimates (up to systematic deviations that depend
on the clusters used) of the error threshold.

\subsection{Zeroth-order approximation}

The effects of the depolarizing channel on topological codes can be
expressed by a spin-glass model with the partition function
\cite{nishimori:02}
\begin{equation}
	Z[A] =
	\sum_{\phi_i}\sum_{\tilde{\phi}_i}\prod_{\langle ij \rangle}
	A^{(\tau^x_{ij},\tau^z_{ij})}_{\phi_i-\phi_j,
		\tilde{\phi}_i-\tilde{\phi}_j},
\end{equation}
where
\begin{equation}
	\begin{split}
		A^{(\tau^x,\tau^z)}_{\phi,\tilde{\phi}} = & \exp\{\beta J\tau^x
		\cos \pi \phi + \\
		& \beta J\tau^z\cos \pi \tilde{\phi} + \beta J\tau^x\tau^z \cos
		\pi\phi \cos\pi\tilde{\phi}\}\,.
	\end{split}
\end{equation}
$\tau^x_{ij} \in \{\pm 1\}$ and $\tau^z_{ij} \in \{ \pm 1\}$ are
quenched random variables chosen from the distribution
\begin{equation}
	P(\tau_{ij}^x,\tau_{ij}^z)
	\propto
	{\rm e}^{\beta J_{p} (\tau^x_{ij} + \tau^z_{ij}
		+ \tau^x_{ij}\tau^z_{ij})}\,.
\end{equation}
This model has a gauge symmetry in the subspace $J=J_p$ which
corresponds to the Nishimori line.

To determine the multicritical point we replicate the partition
function to take into account the quenched randomness of the variables
$\tau^x_{ij}$ and $\tau^z_{ij}$, i.e.,
\begin{equation}
	Z_n[A] = \left[ \left( \sum_{\phi_i}\sum_{\tilde{\phi}_i}
	\prod_{\langle ij \rangle}
	A^{(\tau^x_{ij},\tau^z_{ij})}_{\phi_i-\phi_j,
		\tilde{\phi}_i-\tilde{\phi}_j}
	\right)^n
	\right]_{\rm av} ,
\end{equation}
where the brackets denote a configurational average.
The local Boltzmann factor is then given by
\begin{equation}
	A_{n,k} =
	\left[\prod_{\alpha=1}^n
	A^{(\tau^x_{ij},\tau^z_{ij})}_{\phi^{\alpha}_i-\phi^{\alpha}_j,
		\tilde{\phi}^{\alpha}_i-\tilde{\phi}^{\alpha}_j}
	\right]_{\rm av}
\end{equation}
where $k$ distinguishes the specific configuration
$(\phi_i^{\alpha},\tilde{\phi}_i^{\alpha})$. The $n$-binary Fourier
transformation gives the dual Boltzmann factor $A^{*}_{n,k}$. It follows
\cite{ohzeki:09,ohzeki:09a} that $A_{n,0} = A^*_{n,0}$ determines the
critical point along the Nishimori line. Taking the leading term in $n$,
we obtain the error threshold for the depolarizing channel of the toric
code as $p_c = 0.189\ldots$ under the conditions $J=J_p$ and
$3\exp(-4\beta J)=p/(1-p)$ for the Nishimori line. Because the local
Boltzmann factors for the topological color codes on both the hexagonal
and square-octagonal lattice are the same, we obtain the same estimate
for the error threshold.

\subsection{First-order approximation using finite clusters}

To reduce systematic errors we consider finite-size clusters with four
bonds on each square lattice for the toric code, six triangles taken
from each triangular lattice for the color codes on the hexagonal
lattice and four triangles from each union jack lattice for color codes
on the square-octagonal lattice (see Fig.~\ref{fig:duality}). We
compute the principal Boltzmann factors on the clusters, i.e.,
\begin{equation}
	A^{(1)}_{n,0} = \left[\left(\sum_{\phi_0,\tilde{\phi}_0}\prod_{(ij)}
	A^{(\tau^X_{ij},\tau^Z_{ij})}_{\phi_0.\tilde{\phi_0}} \right)^n
	\right]_{\rm av} ,
\end{equation}
as well as its dual $A^{*(1)}_{n,0}$ via a $n$-binary Fourier
transformation. As before, the critical point along the Nishimori line
is determined via $A^{(1)}_{n,0} = A^{*(1)}_{n,0}$. Taking the leading
order in $n$ we obtain for the error thresholds
\begin{eqnarray}
p_c &=& 0.1888\ldots \,\, {\rm (toric\,\, code)}, \\ 
p_c &=& 0.1914\ldots \,\, {\rm (color\,\, code - hexagonal)}, \\
p_c &=& 0.1878\ldots \,\, {\rm (color\,\, code - sq.\,\, octogonal)}. 
\end{eqnarray}
There are small variations in the estimates, however, the estimates are
all of the order of approximately 19\% and in agreement with our results
from Monte Carlo simulations.

\section{Results and Conclusion}

We have shown that the stability under depolarizing noise of toric codes
can be related to the existence of a magnetic phase in a random eight-vertex
model. Similarly, color codes turn out to be related to a class of
`interacting' eight-vertex models.  We analyze the models resulting from the
mapping via both large-scale parallel tempering Monte Carlo
simulations \cite{katzgraber:09c,andrist:10} and the duality
method \cite{ohzeki:09,ohzeki:09a}. By determining $T_c(p)$ for
different error probabilities $p$, we are able to determine the phase
boundary in the $p\,$---$T$ plane (Fig.~\ref{fig:phasediagram}). Both
approaches confirm the existence of a stable ordered phase and by
locating the intersection of the phase boundary with the Nishimori line,
we compute in a nonperturbative way, the disturbing effects of noise on
topological codes. The external noise considered in this work is more
realistic than in previous studies because it applies to both bit-flip
errors, phase-flip errors and more importantly, a nontrivial combination
thereof.

The error threshold to depolarization errors for different classes of
topological codes studied is approximately $19\%$, which is larger than
the threshold for noncorrelated errors. This is very encouraging and
shows that topological codes are more resilient to depolarization
effects than previously thought. The profound relationship between
complex statistical-mechanical models, such as the eight-vertex model, and
topological quantum error correction promises to deliver a plethora of
new paradigms to be studied in both fields in coming years.

\begin{acknowledgments}

We would like to thank H.~Nishimori and D.~Poulin for useful
discussions.  M.A.M.-D.~and H.B.~thank the Spanish MICINN grant
FIS2009-10061, CAM research consortium QUITEMAD S2009-ESP-1594, European
Commission PICC: FP7 2007-2013, Grant No.~249958, UCM-BS grant
GICC-910758.  Work at the Perimeter Institute is supported by Industry
Canada and Ontario MRI.  H.G.K.~acknowledges support from the Swiss
National Science Foundation (Grant No.~PP002-114713) and the National
Science Foundation (Grant No.~DMR-1151387).  M.O.~acknowledges financial
support from Grant-in-Aid for Young Scientists (B) No.~20740218 by MEXT
and Kyoto University's GCOE Program {\em Knowledge-Circulating Society}.
The authors acknowledge ETH Zurich for CPU time on the Brutus cluster
and the Centro de Supercomputaci{\'o}n y Visualisaci{\'o}n de Madrid
(CeSViMa) for access to the Magerit-2 cluster.

\end{acknowledgments}

\bibliography{refs,comments}

\begin{thebibliography}{40}
\bibitem[{com()}]{comment:ref}
\bibinfo{note}{Upon completion of this work we learned of a new algorithm
  \cite{wootton:12} based on parallel tempering Monte Carlo that attains an
  error threshold of $p_c\simeq 18.5\%$.}
\end{thebibliography}

\end{document}